\documentclass{CSITproc}
\usepackage{amsmath}

\begin{document}

\title{ Recursive matrix algorithms, distributed dynamic control, scaling, stability
 }

\numberofauthors{1} 

\author{
\alignauthor Gennadi Malaschonok \thanks{ Preprint of the paper:   G. Malaschonok, "Recursive Matrix Algorithms, Distributed Dynamic Control, Scaling, Stability," 2019 Computer Science and Information Technologies (CSIT), Yerevan, Armenia, 2019, pp. 112-115, doi: 10.1109/CSITechnol.2019.8895255 } \\ 
    \affiliation{National University of Kyiv-Mohyla Academy \\ Kyiv, Ukraine }
    \email{malaschonok@gmail.com}
}

\date{23 - 27 September Yerevan, Armenia}
\maketitle

\abstract{The report is devoted to the concept of creating block-recursive matrix algorithms for computing on a 
supercomputer with distributed memory and dynamic decentralized control.  
}


\section{Introduction}

Appearance of the supercomputer system with hundreds of thousands of cores poses many new problems
for specialists in the field of parallel computing. The three main ones are uniform load of equipment,
the presence of control over the growth of the error of numbers during calculations and the presence of protection 
against possible physical failures of individual processors.

To ensure uniform loading of equipment, two different approaches are distinguished: static and dynamic.

In the paper [1], the authors presented a new task insertion extension for PaRSEC,
Dynamic Task Discovery (DTD), supporting shared and distributed memory environments. 
They compare two programming paradigms: Parameterized Task Graph (PTG) and Dynamic Task Discovery (DTD).
The result shows good scalability and comparable
result to PTG in most cases and, where comparable benchmarks exist, consistently better performance compared to other runtime.

We propose another dynamic control scheme for a parallel computing process,
which is much simpler than DTD and does not allow to control the parallel execution of an arbitrary algorithm.
It can be used only for block-recursive algorithms.
In such algorithms, independent separate subtasks operations applied to blocks, so it is easy to organize decentralized control of the entire computational process.
We give examples of such algorithms and describe the basic data structures for managing distributed parallel computing.

 The second problem is the accumulation of errors during calculations.
 The larger the matrix size, the more error can accumulate.

Let a set of matrices be given and it is required to calculate some new matrices, vectors or scalars. 
All source numbers are rational numbers due to the fact that the memory has a finite size. 
If your algorithm uses only rational operations, then you have the opportunity to get an exact answer
 with respect to the input data.

If the approximate calculations are used, then  the calculations  error increases with the number of operations. 
Consequently, with the growth of matrix sizes, there comes a moment when the error exceeds the allowed limits. 
For example, in the Gauss algorithm, errors can exceed the exact solution already for matrices 
of order 10 if these matrices are ill-conditioned. Unfortunately,
for every well-conditioned matrix, this boundary also has a well-defined value. 
And what should be done if the size of the matrix exceeds this limit value?

Then you have to change the computational paradigm.
For example, you can exchange accuracy for time, but you have many differen possibilities to do this. 
The question is, what should be the new computational paradigm?
It may vary depending on the type of algorithm. Let's consider these types.

\section{Three classes of matrix algorithms}

All matrix algorithms are divided into three separate classes.

The rational direct matrix algorithms  -- $ MA_1 $.

The irrational direct matrix algorithms -- $ MA_2 $.

The iterative matrix algorithms -- $ MA_3 $ .

The first class ($ MA_1 $) contains algorithms that use only four arithmetic operations. As a result, only rational functions can be computed. This class includes an algorithm for solving systems of linear equations, calculating the inverse matrix,   a determinant,   a similar three-diagonal matrix,   a characteristic polynomial, a generalized inverse matrix, a kernel of a linear operator, LU, LEU and LDU decompositions, Bruhat decomposition and so on.

The second ($ MA_2 $) class consists of all direct methods that did not fall into the first class. Elements of matrices that are obtained as a result of the application of these methods cannot be obtained in the form of rational functions. This class includes algorithms for QR-decomposition of matrices, calculations of a similar two-diagonal matrix, and others.

The third class ($ MA_3 $) consists of all remaining algorithms, in which iterative methods are used. For example, algorithms for calculating eigenvalues and eigenvectors of a matrix and algorithms for SVD decomposition fall into this class when the rank of the matrix is greater than four.

Here you can see a complete analogy with algorithms for solving algebraic equations.
Algorithms for solving algebraic equations can be divided into the same three classes. The first class contains algorithms for solving linear equations. The second class consists of direct algorithms for solving equations of the second, third and fourth degree. And the third class consists of iterative algorithms for solving algebraic equations. Such algorithms allow finding solutions to equations of degree five and higher.

Each of these classes uses its own special matrix algorithms. Accordingly, each of these classes requires its own methods of creating matrix algorithms for large matrices and for supercomputers with distributed memory.

\section{  MA$1$-algorithms}
We will assume that all matrices are square and have $2^k$ rows and columns. If the matrix has other sizes, then it can be added to such a square matrix with zero or unit blocks.
\subsection{Recursive standard and Strassen's matrix multiplication}
 
Recursive algorithm for standard matrix multiplication is based on the equation
$$
\begin{pmatrix} A_0 & A_1 
\\ A_2 & A_3 \end{pmatrix}
\times \begin{pmatrix} B_0 & B_1
\\ B_2 & B_3 \end{pmatrix} +\begin{pmatrix} C_0 & C_1 \\ C_2 & C_3 \end{pmatrix}=
\begin{pmatrix} D_0 & D_1 \\ D_2 & D_3 \end{pmatrix}$$ 
So 
$D_0=A_0 B_0 +    (A_1 B_2+ C_0), 
 D_1=A_0 B_1 +    (A_1 B_3+ C_1), 
 D_2=A_2 B_0 +    (A_3 B_2+ C_2), 
 D_3=A_2 B_1 +    (A_3 B_3+ C_3). $
 
Number of operations for the standard algorithm is $\sim n^3$.

The Strassen multiplication algorithm \cite{Strassen} is also a block recursive algorithm.
The number of operations for this algorithm is $\sim n^{\log_2 7}$.
There exists a boundary with respect to the density of the matrix, which separates the region of applicability of the Strassen multiplication. 
  (see details in \cite{MVL}).

\subsection{ Recursive  inversion of triangular  matrix}
If $\mathcal A=\begin{pmatrix} A & 0 \\ B & C \end{pmatrix}$ is invertible triangular matrix of order $2^k$ and
 $\det({\mathcal A})\neq 0$ then  
 $${  \mathcal A}^{-1}= 
\begin{pmatrix}  A ^{-1} & 0  \\-C^{-1} B A^{-1} & C^{-1}  \end{pmatrix}.
$$
 
\subsection{ Recursive Cholesky decomposition}
 
Let $\mathcal A=\begin{pmatrix} A_1 & A_2 \\ A_2^T & A_3 \end{pmatrix}$ be a positive definite symmetric matrix and 
$H=\begin{pmatrix} B & 0 \\ C & D \end{pmatrix}$ be a low triangle matrix with the property ${\mathcal A}= H H^T$.
The mapping 
$$
  Chol({\mathcal A} )=(H, H^{-1}) 
$$
 is called an {\it Cholesky decomposition}.
It is easy to see that the recursive algorithm of Cholesky decomposition has the following form.
  Let 
  $$Chol({  A_1} )=(B, B^{-1}).$$ 
 
 Then we can compute 
 $$C=A_2^T (B^{-1}) \ \  \hbox{and}  \  \  F=A_3- C C^T $$
 Let 
 $$Chol({F} )=(D, D^{-1}). $$
  
 Then $H=\begin{pmatrix} B & 0 \\ C & D \end{pmatrix}$ and $H^{-1}=\begin{pmatrix} B^{-1} & 0 \\ -D^{-1} C B^{-1} & D^{-1} \end{pmatrix}$.
 
 \subsection{Recursive Strassen's matrix inversion}

If $\mathcal A=\begin{pmatrix} A_0 & A_1 \\ A_2 & A_3 \end{pmatrix}$,
 $\det({\mathcal A})\neq 0$  
 and $\det(A_0)\neq 0$  then the inverse matrix can be calculated as follows \cite{Strassen} 
 $  $
 $$  {\mathcal A}^{-1}= 
\begin{pmatrix}\textbf{\emph{I}} & -A_0^{-1}A_1 \\ 0 & \textbf{\emph{I}} \end{pmatrix}  
\begin{pmatrix} \textbf{\emph{I}} & 0 \\ 0 & (A_3-A_2A_0^{-1}A_1)^{-1} \end{pmatrix} 
 $$
 $$
\times\begin{pmatrix} \textbf{\emph{I}} & 0 \\ -A_2 & \textbf{\emph{I}} \end{pmatrix} 
\begin{pmatrix} A_0^{-1} & 0 \\ 0 & \textbf{\emph{I}} \end{pmatrix}
=\begin{pmatrix}
M_6   & M_1  M_4 \\ M_5   & M_4
\end{pmatrix}.
$$
We have denoted here  $M_0=-A_0^{-1}$,\  $M_1=M_0 A_1$,\  $M_2=A_2 M_0$,\  
$M_3=M_2 A_1$,\  $M_4=(A_3+M_3)^{-1}$,\  $M_5=-M_4  M_2$,  $M_6=M_1  M_5-M_0$.

%

\subsection{Other recursive matrix algorithms of MA$1$-class}

You can find many other recursive matrix algorithms of this class in the papers \cite{14}-\cite{2015}. These are such algorithms as
  computation of the adjoint matrix, kernel and matrix determinant, computation of the
generalized Bruhat decomposition in fields and in commutative domains,  LEU and LDU triangular decomposition of matrices.

New applications of these algorithms were preposed in \cite{2018a} and \cite{2018b}.

As we can see, many block recursive algorithms are already known in the class $MA_1$. 
However, in the next class $MA_2$, we know only one such  algorithm. 
This is Schonhage block-recursive algorithm for the 
QR-decomposition of a matrix \cite{Schonhage}. See also \cite{Tiskin} and \cite{2018A}.

In the next section, we propose another way of presenting algorithm \cite{Schonhage} and we calculate the exact number of operations in
the case of the decomposition of square matrices whose size is equal to the power of the number 2.

\section{ MA$2$-class: QR decomposition}
 
 Let $ A $ be a matrix over a field. It is required to find the upper triangular matrix 
 $ R $ and the orthogonal $ Q $ matrix such that $ A = QR $. For definiteness, we will 
 consider an algorithm applied to a square matrix $ A $ over a field of real numbers.
 
 Consider the case of a $2\times 2$ matrix. The desired decomposition $ A = QR $ has the form:
 $$   \left(\begin{array}{cc}\alpha  & \beta \\  \gamma &  \delta \end{array}\right) =   \left(\begin{array}{cc}c & -s \\ s & c \end{array}\right)    \left(\begin{array}{cc}a  & b \\ 0 & d \end{array}\right) , $$
where the numbers $ s $ and $ c $ satisfy the equation $ s ^ 2 + c ^ 2 = 1 $.

After multiplying from the left of both sides of the equation by the inverse matrix $ Q ^ {- 1} = Q ^ T $, we get: $Q^T A=R$. 
  
  If $ \gamma = 0 $ then we can set $ c = 1, \ s = 0 $. If $ \gamma \ne 0 $, then
  $ \Delta = \alpha ^ 2 + \gamma ^ 2 > 0.  $
   Then we get
 $  c \alpha+ s \gamma=a,  \ \  \ c \gamma -s\alpha =0 $ and  
 $  c= a \alpha /\Delta,  \ \  s= a \gamma/ \Delta.   $
 
 Therefore, $ 1 = s ^ 2 + c ^ 2 = a ^ 2 / \Delta $, hence $ | a | = \sqrt {\Delta} $.
 $  c= \alpha/ {\sqrt \Delta },  \ \ \ \  s=   \gamma/ {\sqrt \Delta }. $
 
 We denote such a matrix $ Q $ by $ g _ {\alpha, \gamma} $.

Let the matrix $ A $ be given, its elements $ (i, j) $ and $ (i + 1, j) $ be $ \alpha $ and $ \gamma $, and all the elements to the left of them be zero: $ \forall (s <j ) :( a_{i, s} = 0) \ \& \ (a_{i + 1, s} = 0) $.

We first describe the well-known sequential algorithm.

  Let $ G_ {i, j} = \, \mathbf{diag} (I_ {i-1}, g _ {\alpha, \gamma}, I_ {n-i-1}) $. These matrices are called Givens matrices. Then the matrix $ G_ {i, j} A $ differs from $ A $ only in two rows $ i $ and $ i + 1 $, but all the elements to the left of the column $ j $ remain zero, and in the column $ j $ in $ i + 1 $ line will be 0.
 
 This property of the Givens matrix allows us to formulate such an algorithm
 $$\bf\textbf{  Sequential algorithm }$$

 \noindent
 (1).  
First we reset the elements under the diagonal in the left column:
 $$ A_{1}=G_{1,1}G_{2,1}... G_{n-2,1}G_{n-1,1}A$$

  \noindent
 (2).  
Then we reset the elements that are under the diagonal in the second column:
 $$ A_{2}=G_{2,2}G_{3,2}... G_{n-2,2}G_{n-1,2}A_{1}$$
 
 \noindent
 (k).   Denote $G_{(k)}= G_{k,k}G_{k-1,k}... G_{n-2,k}G_{n-1,k} $, for $k=1,2,..,n-1$.  
 Then, to calculate the elements of the $ k $ th column, we need to obtain the product of matrices
  $$ A_k=G_{(k)} A_{k-1}.$$
 \noindent (n-1).  
At the end of the calculation, the element in the $ n-1 $ column will be reseted: $A_{n-1}=G_{(n-1)}A_{n-2}=G_{n-1,n-1} A_{n-2}.$

\subsection{ QR$_G $ decomposition}
 Let a matrix $ M $ of size $ 2n \times 2n $ be divided into four equal blocks: 
 $ M = \left (\begin {array} {cc} A & B \\ C & D \end {array} \right) $. 
 There are three stages in this algorithm.
  $$ \bf\textbf{  $QR_G$  algorithm }$$ 

 \noindent (1). The first stage is the $ QR_G $ decomposition of the block  
  $C$:  
  $$C=Q_{1}C_{1},\ M_{1}=\, \mathbf{diag}(I,Q_{1}) M= \left(\begin{array}{cc}A & B \\ C_{1} & D_{1} \end{array}\right).$$

   \noindent
 (2). The second stage is the cancellation of a parallelogram composed of two triangular blocks: the lower triangular part $ A ^ L $ of the block $ A $ and the upper triangular part  $ C ^ U_ {1} $ of the block $ C_ {1} $. Denote the upper triangular matrix $ A_ {1} $ and annihilating matrix $ Q_ {2} $:

 $$ Q_{2} \left(\begin{array}{c}A \\ C_{1} \end{array}\right) = \left(\begin{array}{c}A_{1} \\ 0 \end{array}\right),\   M_{2}= Q_{2} M_{1}= \left(\begin{array}{cc}A_{1} & B_{1} \\ 0 & D_{2} \end{array}\right).$$

  \noindent
 (3). The third stage is the $ QR_G $ decomposition of the $ D_ {2} $ block:
   $D_{2}=Q_{3}D_{3}$.  
   $$R=\, \mathbf{diag}(I,Q_{3}) M_{2}= 
   \left(\begin{array}{cc}A_{1} & B_{1} \\ 0 & D_{3} \end{array}\right). $$

 As a result, we get:
 $$M=Q^T R, \ \ Q=\, \mathbf{diag}(I,Q_{3}) Q_{2} \, \mathbf{diag}(I,Q_{1}). $$
 
 Since the first and third stages are recursive calls of the $ QR_G $ procedures, it remains to describe the parallelogram cancellation procedure. Let's call it a QP decomposition.

\subsection{ QP-decomposition} 
 
Let the matrix $ M = \left (\begin {array} {c} A \\ B ^ U \end {array} \right) $ have dimensions $ 2n \times n $ and, at the same time, the lower unit $ B ^ U $ of size $ n \times n $  has an upper triangular shape - all elements under its main diagonal are zero. We are looking for the factorization of the matrix $ M = QP = Q \left (\begin {array} {c} A ^ U \\ 0 \end {array} \right) $, with the orthogonal matrix $ Q $.

It is required to annul all elements between the upper and lower diagonals of the $ M $ matrix, including the lower diagonal. It is easy to see that this can be done with Givens matrices. We will consistently perform column invalidation by traversing column elements from bottom to top and traversing columns from left to right.

But we are interested in the block procedure. Since $ n $ is even, we can break the parallelogram formed by the diagonals into 4 parts using its two middle lines. We get 4 equal parallelograms. To cancel each of them, we will simply call the parallelogram cancellation procedure 4 times. We will perform the calculations in this order: the bottom left ($ P_ {ld} $), then we simultaneously cancel the top left ($ P_ {lu} $) and the bottom right ($ P_ {rd} $), and last we will cancel the top right ( $ P_ {en} $). The corresponding orthogonal Givens matrices of size $ n \times n $ are denoted $ Q_ {ld} $. $ Q_ {lu} $. $ Q_ {rd} $ and $ Q_ {ru} $.
Let 
 $$ \bar Q_{ld}=  \, \mathbf{diag}(I_{n/2},Q_{ld} ,I_{n/2}),\ \   \bar Q_{ru}= \, \mathbf{diag}(I_{n/2},Q_{ru} ,I_{n/2}),$$
  As a result, we get: 
 $$Q=   \bar Q_{ru}   \, \mathbf{diag}(Q_{lu}  ,Q_{rd} ) \bar  Q_{ld}    $$

 The number of multiplications of matrix blocks of size $ n / 2 \times n / 2 $ is 24. Hence the total number of operations: $ Cp (2n) = 4Cp (n) + 24M (n / 2). $. Suppose that for multiplication of two matrices of size $ n \times n $ you need $ \gamma n ^ \beta $ operations and $ n = 2 ^ k $, then we get:
 $ Cp(2^{k+1}) = 4 Cp(2^{k}) + 24 M(2^{k-1})=4^{k }Cp(2^1)$ $ +24\gamma\sum_{i=0}^{k-1} 4^{k-i-1}  2^{i\beta}= $ \\ 
 $24 \gamma (n^2/4)\frac{2^{k(\beta-2) }-1}{2^{(\beta-2)}-1}+6n^2 $ $=6 \gamma  \frac{n^{\beta }-n^2}{2^{\beta }-4}+6n^2$ 
    $$ Cp(n)  =  \frac{6 \gamma n^{\beta } }{2^\beta(2^{\beta}-4)}+ \frac{3n^2 }{2}(1- \frac{ \gamma  }{  2^{\beta}-4 })$$ 

 6

\subsection{ The complexity of  QR  decomposition algorithm}
 Let us estimate the number of operations $ C (n) $ in this block-recursive decomposition algorithm, 
 assuming that the complexity of the matrix multiplication is $ M (n) = \gamma n ^ \beta $, the complexity of 
 canceling the parallelogram is $ Cp (n) = \alpha  ^\beta $, where $ \alpha, \beta, \gamma $ are constants, 
  $\alpha= \frac{6 \gamma }{2^\beta(2^{\beta}-4)}$ and $n=2^k$:
  $C (n)=2C (n/2) + Cp(n) + 6M(n/2)=$ $2C (2^{k-1})$ $ + Cp(2^{k}) + 6M(2^{k-1})= $
%
%
$$=  \frac{\gamma 6 (2^{\beta  }-3)( n^{\beta }-\frac{2n}{2^\beta} )}{ (2^{\beta}-4)(2^{\beta  }-2)} $$

\section{Dynamic algorithms}


 
Dynamic matrix algorithms are based on matrix block-recursive algorithms. 
In such algorithms, the matrix is recursively divided into blocks. 
A block-recursive algorithm is again applied to each of the blocks. 
This happens as long as the blocks remain large enough. When the block size becomes small enough, 
the usual sequential algorithms are applied to the blocks. 
This limit for the size of a small block depends on the physical characteristics of the computing device
and should be automatically adjusted to the specific equipment.

\subsection{The dynamic algorithm has three stages}

First stage. This is the initial construction of the connections tree for computational nodes. The large blocks are sent from the root node to a child along with lists of free nodes. From these child nodes, data is sent further, but already with smaller blocks and corresponding parts of the list of free nodes.

Second stage. It occurs when either all the free nodes have received their subtasks, or when the size of the blocks has decreased to a certain boundary, which is predetermined. The tree of connections is constructed and the calculation takes place on leaf vertices.

The third stage.  At this stage, the results are returned from leaf vertices to the root vertex. The result of the main task is obtained at root vertex and the calculations are completed.

\subsection{Automatic redistribution of subtasks}

Dynamic control involves the automatic redistribution of subtasks from overloaded nodes to free nodes. For this purpose, a scheme is provided for transmitting information about free nodes and information about overloaded nodes.
Both streams of information are transmitted along the tree towards the root vertex until they meet at a certain node. After this, the information about free vertices is redirected to the overloaded vertices.

The largest subtasks from the overloaded nodes are transmitted to the free nodes. And after completing the calculations, the result is returned to the node from which this subtask was obtained.
 
\subsection{Protection scheme in case of a failure of a node}

It also uses a very simple protection scheme in case of a failure of a node during calculations.

Let node 1 send a subtask S to node 2. Let node 2 fail and the failure message came to node 1. Node 1 will mark this subtask S as unsolved and return it to the list of unsolved subtasks. All operations of transferring results from child nodes to node 2 are simply canceled. No other action is required. The computational process will continue on all other nodes without any changes.

Note that such a protection scheme has significant advantages over the protection scheme in static algorithms.

\section{Distributed dynamic control mechanism for computational process }
	Consider the components of the control mechanism of the computational process.

	The two main objects are Drop and Amine. We assign Amine to the task, and we call  Drops the components of its subtasks. In the calculation of Drop, it first creates his Amin, which consists of the Drops for the next recursion.
	
\subsection{Drop}
	We divide the computational graph into separate compact subgraphs (Drops).

For example, consider  recursive  inversion of triangular  matrix.
If $\mathcal A$  is invertible triangular matrix  then 
 $$
 {\mathcal A}=\begin{pmatrix} A & 0 \\ B & C \end{pmatrix}, \ \ \
 {  \mathcal A}^{-1}= 
\begin{pmatrix}  A ^{-1} & 0  \\-C^{-1} B A^{-1} & C^{-1}  \end{pmatrix}.
$$
We have here two Drops for triangular matrix inversion  $  G=C^{-1}$ and  $  F=A^{-1}$, one Drop for matrix multiplication $H=B\times  F$  one Drop for matrix multiplication with change the signs of elements  $-G\times H$.

Thus, we define the Drops as the smallest components of the computational graph that can be transferred to other processors. 

\subsubsection{The main fields of the Drop object} \

\noindent --- PAD   --- address of this Drop  (the nunber of processor, the nunber of Amine, the nunber of this Drop in its Amine).

\noindent --- Type --- Drop type (unique number in the list of all Drop types).

\noindent --- InData and OutData --- these are vectors for input and output data.

\noindent --- Amine --- the Amine of this Drop.

\noindent --- RecNum --- recursion number of this Drop.

\noindent --- Arcs ---  graph topology of Amine (the connection topology diagram for data transfer during calculations).

\subsection{Amine}
Before the Drop is calculated, we need to expand the corresponding subgraph. This subgraph is called Amine. This Amine also consists of Drops.

As we see, the Amine of recursive  inversion of triangular  matrix has 4 Drops.
The Amine $A\cdot B$ consists of 4 Drops $A\cdot B$ and 
4 Drops $A\cdot B + C$. And so on.

\subsubsection{The main fields of the Amine object} \

\noindent --- PAD  --- address to return the result of the calculation of this Drop.

\noindent --- Type, inData, outData --- the same as the Drop.

\noindent --- Drop --- an array of all Drops of a given Amine.

\subsection{Pine}
All Amines that are formed in one processor are stored in the general list, which is called Pine. 

\subsection{Vokzal}
At the Vokzal are all the Drop-tasks that are awaiting their direction to the calculations. These tasks are located at different levels. These levels correspond to the depth of recursion for Drops. 
 
\subsection{Aerodrome}
	Each processor that sent a Drop task is called a parent. The list of all parent processors is called an Aerodrome.

\subsection{Terminal}
	The terminal is used to communicate with the child processors that were sent Drop-tasks. All child processors are stored in the terminal.

\subsection{Two computational threads}

We use two threads: a computational thread and a dispatcher thread. These threads will run on each cluster processor. 

\subsubsection{CalcThread}

The CalcThread waits for the arrival of the first Drop task at the vokzal and starts the corresponding calculations.

{\bf	CalcThread objects:} \

\noindent --- Pine --- list of Amines on this processor.

\noindent --- Vokzal --- an array of lists of available Drop tasks.

\noindent --- Aerodrome --- list of parent processors.

\noindent --- Terminal --- an array of child processor lists.

\noindent --- CurrentDrop --- current Drop, which is calculated.

{\bf  CalcThread functions:} \

\noindent --- WriteResultsToAmin --- the results of a Drop calculation are written to its Amine in the input data vectors of other Drops.

\noindent --- InputDataToAmin --- create an Amine from a Drop, if a new task arrives, we make an input function.

\noindent --- WriteResultsAfterInpFunc --- write the result of the input function to the all Drops.

 \noindent  ---  runCalcThread --- main function of CalcThread.

\subsubsection{Dispatching Thread}

The work of the dispatching thread can be divided into 10 processes:

\noindent --- Waiting for completion signal.

\noindent --- Reception task.

\noindent --- Receive free processors.

\noindent --- Receive and record the status of the child processor.

\noindent --- Receive the result of the calculated Drop and record these results in the corresponding Amine.

\noindent --- Receive non-main components and record it in the right place.

\noindent --- Sending available tasks to free processors (if there are tasks and processors).

\noindent --- Sending the list of free processors to a child (if there are no Drop tasks available, but there are the list of free processors and child processors).

\noindent --- Sending the entire list of free processors to the parent processor (if the Vokzal is empty and the Terminal does not contain overloaded child processors).

\noindent --- Sending Drop results to parent processors.

\noindent --- Sending non-main components to child processors.

\bigskip

This scheme was implemented in the Java programming language using the OpenMPI and MathPartner \cite{2017} packages, and its work was tested on the matrix multiplication and matrix inversion algorithms.

 A more detailed description of this scheme is presented in \cite{MS2018}.

\section*{Conclusion}

We  proposed a new classification of matrix computational algorithms,
which decomposes all algorithms into three classes: rational, irrational and iterative. 
We discribed the new computational paradigm: using of the block-recursive matrix algorithms for
creating parallel programs that are designed for supercomputers with distributed memory and 
dynamic decentralized control of the computational process. We have shown many examples of such algorithms.
We proposed a dynamic decentralized computation control scheme.



\end{document}